\documentclass[doublecol]{epl2} 
\usepackage{bm}
\usepackage{amsmath}
\usepackage{amssymb}

\newcommand{\eff}{\mbox{\scriptsize eff}}

\title{Slow dynamics in random media: Crossover from glass to localization transition}

\author{K. Kim\inst{1} \and K. Miyazaki\inst{2} \and S. Saito\inst{1}}
\shortauthor{K. Kim \etal}

\institute{                    
  \inst{1} Institute for Molecular Science - Okazaki 444-8585, Japan\\
  \inst{2} Institute of Physics, University of Tsukuba - Tsukuba 305-8571, Japan
}

\abstract{
We study slow dynamics of particles moving in 
a matrix of immobile obstacles using molecular dynamics simulations.
The glass transition point decreases drastically as the obstacle density increases.
At higher obstacle densities, the dynamics of mobile particles
changes qualitatively from glass-like to a Lorentz-gas-like relaxation.
This crossover is studied by density correlation functions, 
nonergodic parameters, mean square displacement, 
and nonlinear dynamic susceptibility.
Our finding is qualitatively consistent with the results of recent 
numerical and theoretical studies on various spatially heterogeneous systems.
Furthermore, we show that slow dynamics is surprisingly 
rich and sensitive to obstacle configurations.
Especially, the reentrant transition is observed for a particular configuration,
although its origin is 
not directly linked to the similar prediction
based on mode-coupling theory. 
}

\begin{document}

\maketitle

Transport phenomena in inhomogeneous systems 
are of great importance in physics, chemistry, biology, and
engineering~\cite{Havlin2002Diffusion}. 
Many biological systems and composite materials consist of components
of various sizes. 
The interplay between the broad range of length and time scales is 
essential to their dynamical properties.
Recently, slow dynamics and the glass transition  
of such systems has attracted much attention. 
One reason is that it is interesting to understand the effect of the generic
disorder of inhomogeneous systems on dynamic arrest~\cite{Alcoutlabi2005Effects}. 
Another reason is the extremely rich phenomenology that such systems
exhibit near the glass transition point. 
For example, the colloidal suspensions with the short-range attractive
potentials often show a crossover from the glass transition at high
densities to gelation at lower densities, which is triggered 
by the spatial disorder due to aggregation of the colloidal
particles~\cite{Trappe2004Colloidal,Zaccarelli2007Colloidal,Zaccarelli2005Model,Zaccarelli2006Gel}.
Other examples include peculiar glassy dynamics 
in binary colloidal mixtures with disparate size
ratios~\cite{Imhof1995Experimental,
Dinsmore1995Phase,Moreno2006Relaxation},
star polymer mixtures~\cite{Mayer2009Multiple},
or anomalous ion transport in silicate glasses~\cite{Horbach2001Structural,
Horbach2002Dynamics, Voigtmann2006Slow}.
However, the presence of multiple length/time scales have hampered
elucidation of the origin of these interesting behaviors.
For this reason, it is desired to study a simple model system where inherent complexities
are pruned down as much as possible.
The possibly simplest model is a mixture of mobile and immobile
spherical particles.
This system is a minimal model of spatially heterogeneous systems such as
a fluid absorbed in porous media.
It can also be seen as a model of a binary mixture where the
characteristic time scales of constituent particles of each component are well separated.
The slow dynamics of this model is interesting in its own right; 
in the dilute limit of immobile particles, dynamic arrest or the glass
transition takes place at finite densities of mobile particles.
The opposite limit where a single mobile particle moves in a matrix of immobile particles
is a classic Lorentz gas model. 
In this case, the mobile particle undergoes a localization transition at a
finite immobile particle density, whose origin is 
purely geometric~\cite{VanBeijeren1982Transport, Stauffer1994Introduction}.
Recently, Krakoviack studied this model theoretically
for arbitrary densities of mobile/immobile components by combining the
replica method with mode-coupling theory (replica MCT or
RMCT)~\cite{Krakoviack2005LiquidGlass, Krakoviack2005Liquidndashglass,
Krakoviack2007Modecoupling}.
Two important predictions have been made in his studies. 
One is the crossover of dynamics from the glass to localization transition.
When the volume fraction of mobile particles, $\phi_{m}$, is large and
the volume fraction of immobile particles, $\phi_{i}$, is small, the
system undergoes a glass transition, where the onset of slow dynamics is 
signaled by the {\it discontinuous} emergence 
of two-step relaxation, over all wavelengths, in the density correlation functions. 
This transition is referred to as Type $B$ transition in the
glass transition community~\cite{GotzeinHansen1991Liquids}.
As $\phi_{i}$ increases, the glass transition point $\phi_{m}^g$
decreases drastically.
At even larger $\phi_{i}$, dynamics near the transition point
qualitatively changes; one-step slow relaxation sets in at large
wavelengths, where the amplitude of the relaxation curve grows continuously 
and progressively propagates toward the shorter wavelengths as
$\phi_{m}$ increases. 
This dynamics is called Type $A$.
The crossover from Type $B$ to $A$ is especially interesting since 
it is ubiquitous in many heterogeneous systems such as colloidal
gels~\cite{Zaccarelli2006Gel,
Zaccarelli2007Colloidal} and systems with disparate size
ratios~\cite{Moreno2006Relaxation}.
Another prediction of RMCT is the existence of the reentrant
pocket in the low $\phi_{m}$ and high $\phi_{i}$ region, in which
the dynamics of mobile particles is {\it accelerated} as $\phi_{i}$ {\it increases}.
This counter-intuitive result was rationalized as being caused by 
dephasing of the dynamics of the caged particles due to multiple collisions
with other mobile particles~\cite{Krakoviack2005LiquidGlass,
Krakoviack2005Liquidndashglass, Krakoviack2007Modecoupling}.

\begin{figure}[t]
\onefigure[width=.35\textwidth]{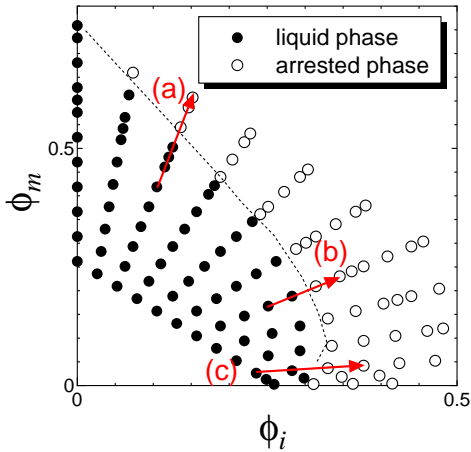}
\caption{
Dynamic phase diagram of a bidisperse soft-sphere system generated by
the EM protocol.
The dotted line represents the Liquid-Arrested (LA), line which is
 defined as the points beyond which the relaxation time $\tau_\alpha$ exceeds $10^3$.
}
\label{3D_soft_diagram}
\end{figure}

In this Letter, we investigate slow dynamics of mixtures of mobile/immobile 
particles by extensive numerical simulations. 
Our goals are twofold; 
first, we study the crossover from the glass to localization 
transition, exploring the whole range of parameter space of $(\phi_{i}, \phi_{m})$. 
Slow dynamics of mobile particles near the transition line is analyzed
using the density correlation functions, 
nonergodic parameters, and mean square displacement.
We also calculate the dynamic susceptibility in order to probe the dynamic
heterogeneities around this crossover. 
The second goal is to study the sensitivity of the dynamics of mobile particles
to a geometry of random configurations of the immobile particles.
To see this, we investigate two types of protocols to generate random configurations.
In the first protocols which we refer to as {\it Quenched-Annealed} (QA),  
we prepare the immobile particles, let them run for some time steps
and, after they are equilibrated, we quench their motions. 
Then, the mobile particles are inserted into the void spaces of the
matrix of immobile particles. 
This QA protocol is appropriate to study dynamics in a porous medium. 
This is the protocol that has been adopted  in RMCT
analysis~\cite{Krakoviack2005LiquidGlass,
Krakoviack2005Liquidndashglass, Krakoviack2007Modecoupling}. 
In the second protocol which we call 
{\it Equilibrated Mixture} (EM), we first run all of
particles until they are equilibrated and then freeze the motion of a fraction 
of the particles while keeping the others mobile.
We design this protocol in order to mimic binary mixtures with disparate time
scales~\cite{Imhof1995Experimental, Dinsmore1995Phase,
Moreno2006Relaxation, Voigtmann2008Double} and to study the dynamics 
of the faster (lighter) component, which would be analogous to that of mobile
particles in an EM system.
In our study, we observe no sign of the reentrant transition for QA
systems, which is contrary to the RMCT prediction~\cite{Krakoviack2005LiquidGlass,
Krakoviack2005Liquidndashglass, Krakoviack2007Modecoupling}.  
Surprisingly, however, we do find the reentrant transition for EM systems. 
The physical origin of this reentrance is purely geometric and 
no direct connection with the RMCT prediction is inferred.  

We carried out molecular dynamics (MD) simulations for two systems. 
The first one consists of an equal number of two types of particles. 
The reason to use the bidisperse system was to avoid crystallization 
in large $\phi_m$ and small $\phi_i$ regions.
Our system is composed of $N = N_1 + N_2 = 1000$ particles in a
cubic box with dimension $L=V^{1/3}=10.8$ under periodic boundary
conditions (PBC).
They interact via the soft potential
$v_{ab}(r) = \epsilon(\sigma_{ab} /r)^{12}$, where $\sigma_{ab} =
(\sigma_a + \sigma_b)/2$ and $a, b \in \{1, 2\}$.
The size and mass ratio were $\sigma_2/\sigma_1 = 1.2$ and 
$m_2/ m_1= 2$, respectively.
The units of length, time, and temperature were taken as $\sigma_1$,
$\sqrt{m_1 \sigma_1^2 / \epsilon}$, and $\epsilon / k_B$,
respectively.
$N_i$ ($N_m$) immobile (mobile) particles were chosen from the
$N_1 + N_2$ particles for each simulation run.  
We used the volume fractions defined by $\phi_i=\pi N_i\sigma_{\eff}^3/6V$ and
$\phi_m=\pi N_m\sigma_{\eff}^3/6V$ as the system parameters, where 
$\sigma_{\eff} =T^{-1/12}(\sum_{a,b}\sigma_{ab}^3/4)^{1/3}$ is the
effective diameter of particles~\cite{Yamamoto1998Dynamics}. 
The states investigated here were $\sigma_{\eff}^3=0.5$, $0.6$, $0.7$,
$0.8$, $0.9$, $1.0$, $1.1$, $1.15$, $1.2$, $1.3$, $1.4$, and $1.45$.
The corresponding temperatures were $T=21.61$, $10.42$, $5.624$, $3.297$,
$2.058$, $1.350$, $0.992$, $0.772$, $0.651$, $0.473$,
$0.352$, and $0.306$, respectively.
We controlled $\phi_i$ and $\phi_m$ by changing $N_i$, $N_m$, and $\sigma_{\eff}$. 
The minimum number of mobile particles was chosen as $N_m=10$.
The time steps used were $0.001\sim 0.005$.
The second system we investigated is the monatomic hard spheres.
This system was employed in order to clarify the origin of the configuration
dependence of the dynamics without being obscured by the bidispersity and 
softness of the continuous potential.  
The system includes $N=1000$ hard spheres with mass $m$
and diameter $\sigma$ 
in a cubic box of volume $V$ under PBC.
$\sigma$ and $\sqrt{{m\sigma^2}/{k_BT}}$ were used as the units of length
and time, respectively.
The temperature was fixed as $k_BT=1$. 
The volume fractions $\phi_i$ and $\phi_m$ were controlled by changing 
$N_i$, $N_m$, and $V$. 
For both systems, we carefully checked the sample dependence of the
observables throughout the study.

We first performed simulations of EM systems of the bidisperse soft-sphere system to 
observe the dynamics for the whole $(\phi_{i}, \phi_{m})$-space.
In Fig.~\ref{3D_soft_diagram}, the dynamic phase diagram is plotted as a
function of $\phi_{i}$ and $\phi_{m}$. 
This is drawn by calculating the self part of the intermediate scattering function (ISF) 
for mobile particles defined by 
$F_s(k,t)$ $=$ $N_{m}^{-1}$ $\langle$ 
$\sum_{i=1}^{N_{m}}$ $\mbox{e}^{i\vec{k}\cdot(\vec{r}_i(t)-\vec{r}_i(0))}\rangle$. 
Here $\vec{k}$ is the wavevector, $k=|\vec{k}|$, and 
$\vec{r}_i(t)$ is the position of the $i$-th particle.
We define  the Liquid-Arrest (LA) line that delineates the
liquid and arrested (glass or localization) phase in
Fig.~\ref{3D_soft_diagram} as the points at which 
the structural relaxation time $\tau_\alpha$ 
reaches $10^3$.
Here $\tau_\alpha$ is defined by $F_s(k=2\pi, \tau_\alpha)=0.1$.
It is seen that, in the small $\phi_{i}$ regime, the LA line (or the glass
transition points) drastically decreases as $\phi_{i}$ increases.
This behavior is in accordance with the results of
simulations~\cite{Kim2003Effects, Chang2004Diffusion, Sung2008The, Mittal2006Using} 
and RMCT~\cite{Krakoviack2005LiquidGlass, Krakoviack2005Liquidndashglass,
Krakoviack2007Modecoupling}.
This trend is sustained up to the small $\phi_{m}$ regime, beyond which a small 
reentrant pocket is observed. 
We shall discuss this reentrance later in detail.

\begin{figure}[t]
\onefigure[width=.43\textwidth]{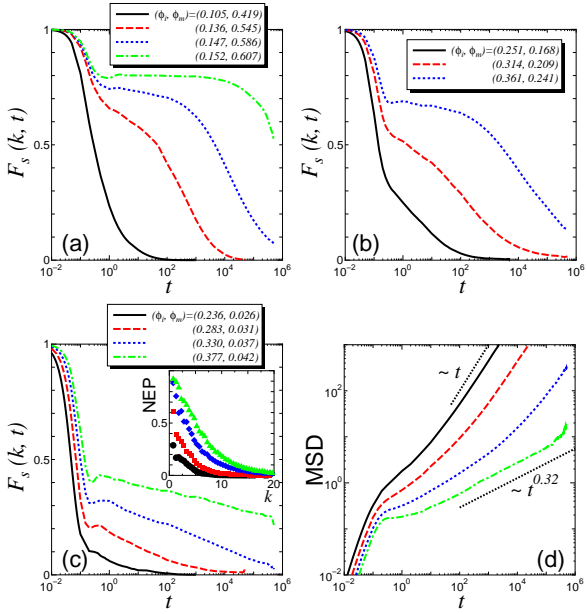}
\caption{
(a)--(c): The self part of the intermediate scattering function $F_s(k, t)$ at $k=2\pi$. 
(a)--(c) corresponds to the cuts indicated in Fig.~\ref{3D_soft_diagram}.
From left to right, the densities $(\phi_i,\phi_m)$ increase along the arrows 
denoted in Fig.~\ref{3D_soft_diagram}. 
Inset of (c): $k$-dependence of the nonergodic parameter.
(d): Mean square displacement for the cut (c) in Fig.~\ref{3D_soft_diagram}.
From top to bottom, the densities $(\phi_i,\phi_m)$ increase.
}
\label{fkt}
\end{figure}

In Fig.~\ref{fkt} (a)--(c), we show $F_s(k, t)$ 
for the mobile particles along the three cuts across the LA line, as
marked by arrows (a)--(c) in Fig.~\ref{3D_soft_diagram}. 
We also calculated the collective part of the ISF, $F(k, t)$ 
and found its behavior is qualitatively the same as $F_s(k,t)$ except at
small wavevectors, the details of which will be explained elsewhere~\cite{Kim2009unpublished}.
Figure~\ref{fkt} clearly demonstrates the existence of two types of distinct dynamics 
depending on the immobile particle densities.
As shown in Fig.~\ref{fkt}(a), at small $\phi_{i}$, $F_s(k,t)$
exhibits a typical glassy behavior; the two-step relaxation with a plateau. 
This plateau or the nonergodic parameter (NEP) 
appears {\it discontinuously} over the whole wavelength range at an onset
density, followed by a mild increase of the plateau heights as the
densities $(\phi_i,\phi_m)$ increase along the arrow.  
This is typical Type $B$ behavior.
At large $\phi_{i}$, on the other hand, 
the ISFs show a single step relaxation with a long tail. Its amplitude or NEP
continuously grows as $(\phi_i,\phi_m)$ increases across the LA line,
as seen in Fig.~\ref{fkt}(c).
The $k$-dependence of the NEP is plotted in the inset of Fig.~\ref{fkt}(c). 
NEP (calculated from $F(k,t)/F(k,0)$) emerges from zero {\it continuously} at small $k$, as 
$(\phi_i,\phi_m)$  approach the LA line.
As $(\phi_i,\phi_m)$ increase further, NEP grows and 
this trend propagates progressively toward larger $k$.
These behaviors are the hallmark of Type $A$ or the localization transition predicted 
by RMCT and also demonstrated by simulations for various spatially 
heterogeneous systems such as colloidal gels~\cite{Zaccarelli2006Gel, Moreno2006Relaxation}. 
The behavior of the mean square displacement (MSD) defined by 
$\delta {r}^2(t)= N_{m}^{-1}\langle \sum_{i=1}^{N_{m}} \Delta\vec{r}_i^2(t) \rangle$,
where $\Delta\vec{r}_i(t)=\vec{r}_i(t)-\vec{r}_i(0)$, 
also qualitatively changes as the dynamics changes from Type $B$ to $A$.
It is well known that, in the Type $B$ regime, MSD shows a plateau caused
by the cage effect (the results not shown).
On the other hand, at the low $\phi_{m}$ limit, 
where the system becomes a Lorentz gas, the anomalous subdiffusive behavior 
is observed at long times, which can be explained in terms of percolation
theory~\cite{Stauffer1994Introduction}.
The latter behavior is observed for the MSD in the Type $A$ regime,
as shown in Fig.~\ref{fkt}(d).
The MSD shows regular diffusive behavior in the liquid region, 
but near the LA line around $(\phi_{i}, \phi_{m}) \approx (0.377, 0.042)$, 
it becomes subdiffusive; 
$\delta {r}^2(t) \approx t^{0.3}$, the exponent of
which is consistent with  $0.32$ predicted for a Lorentz
gas~\cite{Hofling2006Localization, Hofling2007Crossover}.  

\begin{figure}
\onefigure[width=.35\textwidth]{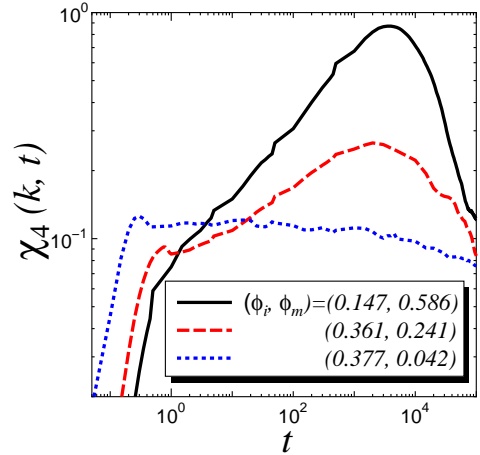}
\caption{
Dynamic susceptibility $\chi_4(k, t)$ at $k=2\pi$ for three
 states on each cut (a)--(c) indicated in
 Fig.~\ref{3D_soft_diagram}(a)--(c) from top to bottom.
}
\label{chi4}
\end{figure}

We next investigated the nonlinear dynamic susceptibility or 4-point 
correlation function, $\chi_4(k, t)$, for mobile
particles~\cite{Glotzer2000Timedependent,Toninelli2005Dynamical,Biroli2006Inhomogeneous}, 
which is a good measure of the dynamic heterogeneities.
$\chi_4(k, t)$ is defined as the variance of the 
fluctuations of the self part of the ISF by 
$\chi_4(k, t) = N_{m} [ \langle \hat F_s(k, t)^2 \rangle-  \langle
\hat{F_s}(k, t)\rangle ^2 ]$.
Here $F_s(k, t)=\langle\hat F_s(k, t)\rangle$.  
In the literature~\cite{Toninelli2005Dynamical}, 
$N_{m}^{-1}\sum_{i=1}^{N_{m}}\cos(\vec{k}\cdot\Delta\vec{r}_i(t))$ is
used as the definition of $\hat F_s(k,t)$. 
With this definition, however, $\chi_4(k, t)$ decays to $1/2$ at $t\rightarrow \infty$.
On the other hand, as we demonstrate here,
the peak of the dynamic susceptibility for the Type $A$ regime grows more mildly than for
the Type $B$ regime. 
In order to demonstrate this suppression of the peak and thus the dynamic heterogeneities 
in the Type $A$ regime without being obscured by a constant plateau at large $t$,  
we have adopted an alternative definition of $\hat F_s(k, t)$ 
which is $N_{m}^{-1}\sum_{i=1}^{N_{m}}\sin(k|\Delta\vec{r}_i(t)|)/k|\Delta\vec{r}_i(t)|$.
Of course, both definitions of $\hat F_s(k,t)$ lead to the identical 
averaged value $F_s(k,t)=\langle \hat F_s(k,t)\rangle$. 
With this new definition, $\chi_4(k, t)$ decays to 0 at  $t \to\infty$.
In Fig.~\ref{chi4}, the time evolution of $\chi_4(k, t)$ is plotted for
the three states (a)--(c) in Fig.~\ref{3D_soft_diagram}.
In the glass (Type $B$) regime at (a),  $\chi_4(k, t)$ exhibits behavior 
well-known for the bulk glass, {\it i.e.}, a pronounced peak at the $\alpha$-relaxation 
time, whose height grows rapidly as the density increases, 
preceded by algebraic growth in the $\beta$-relaxation regime. 
In the localization (Type $A$) regime at (c), on the other hand, 
$\chi_4(k, t)$ does not grow nor show a strong peak, even at long times. 
This implies that dynamic heterogeneities play a minor role in the slow dynamics
in this regime.
A similar behavior of $\chi_4(k, t)$ has been reported for colloidal gels
whose slow dynamics is caused by geometrical
frustrations~\cite{Abete2007Static, Fierro2008Dynamical}.

\begin{figure}
\onefigure[width=.35\textwidth]{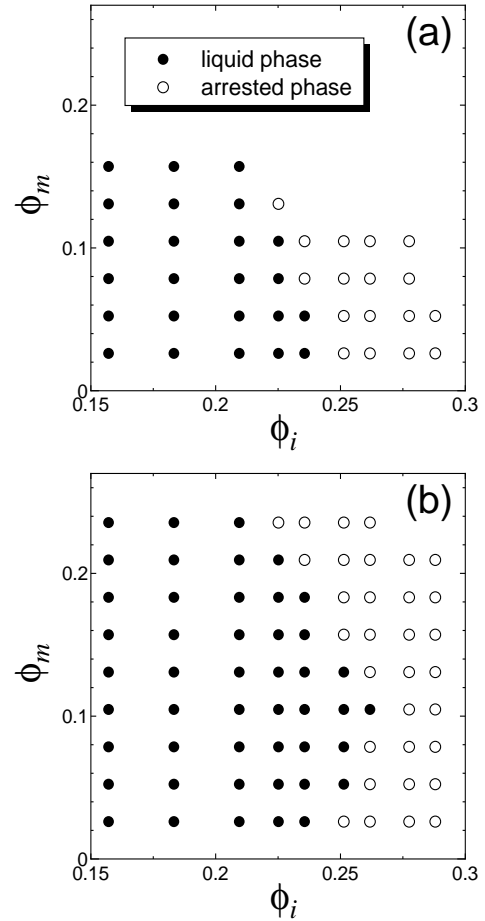}
\caption{
Dynamic phase diagram of hard sphere systems with 
a random matrix generated with two different protocols; (a) QA and (b)
EM protocols.
}
\label{3d_hard_diagram}
\end{figure}

Finally, we examine the sensitivity of the above results to the
configurations of immobile particles by comparing results for the Equilibrated Mixture (EM)
with those using the Quenched-Annealed (QA) protocol.
Hereafter, we mainly focus on the moderately small $\phi_{m}$ regime, because 
little protocol dependence is observed at large $\phi_{m}$. 
Here we switch the system to the monatomic hard sphere system to keep physical
interpretation from being obscured by the bidispersity and continuous
potential of the previous system. 
We checked that the results remain unchanged for the bidisperse
soft-sphere systems.
Dynamic phase diagrams of both QA and EM systems are plotted in Fig.~\ref{3d_hard_diagram}.
Here the LA line is determined as the points beyond which the MSD does
not exceed $10^2$ in a time within the simulation time $t=10^4$.
First, from the simulation for the lowest $\phi_{m}$ of our study ($\phi_{m}=0.026$),  
we estimate the percolation point to be $\phi_{i}^p \approx
0.25$ for both QA and EM systems, which is close to $0.24$ 
determined by Voronoi tessellation~\cite{Sung2008The}.
In the proximity of $\phi_{i}^{p}$, the MSD again shows subdiffusive behavior 
$\delta {r}^2(t) \approx t^{0.3}$~\cite{Hofling2006Localization, Hofling2007Crossover}.
Details of these results will be explained elsewhere~\cite{Kim2009unpublished}.
We here note that to check the finite size effects we performed the same
simulations for $N=10000$ and confirmed that overall behavior is almost
the same in both the QA and EM systems.

As seen in Fig.~\ref{3d_hard_diagram}(a), {\it no} reentrant transition
is observed for QA systems, contrary to the RMCT prediction. 
The LA line monotonically decreases as $\phi_i$ increases, which is compatible with the 
recent numerical simulations for the related QA system~\cite{Single2009Kurzidim}.
This result is hardly surprising, considering that the slow-down of
mobile particle dynamics is mainly due to the geometrical confinement by 
the immobile particles.
Krakoviack has conjectured that 
the reentrant pocket predicted by RMCT is caused
by the dephasing of the trajectories of the caged mobile particles by
collision with other mobile particles~\cite{Krakoviack2005LiquidGlass,
Krakoviack2005Liquidndashglass, Krakoviack2007Modecoupling}.
This effect may be too small to manifest itself as reentrance, if any. 
On the other hand, surprisingly, the EM system has the reentrant pocket,
as clearly seen in Fig~\ref{3d_hard_diagram}(b) and also in Fig.1.
In other words, the dynamics of the mobile particles is {\it accelerated} as 
the number of mobile particles {\it increases}.
The physical origin of the reentrant pocket for EM systems is clear.
The random configuration of the immobile particles is generated 
by quenching their motion after both mobile/immobile particles are 
equilibrated together.
Therefore immobile particles adjust themselves so that the free volume of both
components is entropically maximized and leaves more 
available geometrical pathways for mobile particles, which 
pushes the percolation point $\phi_{i}^{p}$ to larger values and thus
results in faster dynamics of mobile particles. 
Sensitivity of the percolation point on the protocols used to generate
the configuration has been studied in several
contexts~\cite{Chang2004Diffusion,Sung2008The,Mittal2006Using}.
However, its interplay with the dynamics of mobile particles, 
especially the reentrant transition, has not been explored, to the best
of our knowledge. 
It is worth mentioning that an analogous reentrant behavior has been found 
for a distinct system, {\it i.e.}, a binary mixture consisting of large/small 
particles studied by MD simulations~\cite{Voigtmann2008Double}. 
It has been observed that diffusion of small particles 
becomes {\it slower} when the interactions between small particles are 
switched off~\cite{Voigtmann2008Double}.
We believe that this counter-intuitive observation is closely related to
our finding of the reentrant pocket for EM systems 
this switch-off leaves small particles ``invisible'' from each
other and the large ones feel as if there are fewer small
particles around them.
Therefore, the equilibrium configurations of the large (thus less mobile) 
particles would leave less void spaces and pathways for small particles
than before the switch-off. 
This consideration and the results of the present study 
naturally lead to a speculation that 
separation of time scales, rather than disparity of particle sizes,  
is a more essential element behind anomalous dynamics (such as
reentrance) in heterogeneous systems.

In summary, we have carried out MD simulations to study slow dynamics in random media 
for a broad range of parameter space.
Two types of dynamics, $A$ and $B$,  have been observed and
quantified in terms of various dynamic quantities,
which supports RMCT predictions qualitatively.
The dynamic susceptibility clearly demonstrates the minor roles of dynamic
heterogeneities in the Type $A$ regime.
We did not observe any evidence of the reentrant transition for the
system for which RMCT has been originally applied. 
Contrary, we do observe the reentrant transition for the system 
with configurations generated with a different protocol. 
However, the physical origin of this reentrance is not directly
linked to the similar prediction based on RMCT.
Our study suggests that anomalous dynamics common in the heterogeneous
systems might be, more often than not, explained well by a maximally
simplified model such as the one in the present study.

\acknowledgments
The authors thank R. Yamamoto and H. Nakanishi for fruitful discussions.
This work is partially supported by KAKENHI; \# 19740263 (KK),
\# 21740317 (KK), \# 19540432 (KM), \# 21540416 (KM),  and Priority
Areas ``Molecular Theory for Real Systems'' 
(KK, SS) and ``Soft Matter Physics'' (KM).
The computations were performed at Research Center for Computational
Science, Okazaki, Japan.


\end{document}